\documentclass[nofootinbib,twocolumn,superscriptaddress]{revtex4}

\usepackage{graphicx}
\usepackage{bm}
\usepackage{subfigure}
\usepackage{epsfig}
\usepackage{color}
\usepackage{mathtools}
\usepackage{cases}
\usepackage{booktabs}
\usepackage{amsmath}
\usepackage{url}

\usepackage[
colorlinks=true,
filecolor=black,
anchorcolor=blue,
linkcolor=blue,
citecolor=cyan, 
urlcolor=cyan,
linktocpage=true,
plainpages=false,
breaklinks=true,
pdfstartview=FitH
]{hyperref}

\newcommand{\be}{\begin{equation}}
\newcommand{\ee}{\end{equation}}
\newcommand{\bea}{\begin{equation}\begin{aligned}}
\newcommand{\eea}{\end{aligned}\end{equation}}

\newcommand{\ql}{Q_{l}}

\begin{document}

\title{Calibration of the Cryogenic Measurement System of a Resonant Haloscope Cavity}


\author{Dong He}
\affiliation{Key Laboratory of Low-Dimensional Quantum Structures and Quantum Control of Ministry of Education, Key Laboratory for Matter Microstructure and Function of Hunan Province, Department of Physics and Synergetic Innovation Center for Quantum Effects and Applications, Hunan Normal University, Changsha 410081, China}
\author{Jie Fan}
\affiliation{Institute of Physics, Chinese Academy of Sciences, Beijing, 100190, China}
\author{Xin Gao}
\affiliation{College of Physics, Sichuan University, Chengdu 610065, China}
\author{Yu Gao}
\affiliation{Key Laboratory of Particle Astrophysics, Institute of High Energy Physics, Chinese Academy of Sciences, Beijing 100049, China}
\author{Nick Houston}
\affiliation{Institute of Theoretical Physics, Faculty of Science, Beijing University of Technology, Beijing 100124, China}
\author{Zhongqing Ji}
\affiliation{Institute of Physics, Chinese Academy of Sciences, Beijing, 100190, China}
\author{Yirong Jin}
\affiliation{Beijing Academy of Quantum Information Sciences, Beijing 100193, China}
\author{Chuang Li}
\affiliation{College of Mechanical and Electrical Engineering, Wuyi University, Nanping 354300, China}
\author{Jinmian Li}
\affiliation{College of Physics, Sichuan University, Chengdu 610065, China}
\author{Tianjun Li}
\affiliation{CAS Key Laboratory of Theoretical Physics, Institute of Theoretical Physics, Chinese Academy of Sciences, Beijing 100190, China}
\affiliation{School of Physical Sciences, University of Chinese Academy of Sciences, No. 19A Yuquan Road, Beijing 100049, China}
\author{Shi-hang Liu}
\affiliation{Key Laboratory of Low-Dimensional Quantum Structures and Quantum Control of Ministry of Education, Key Laboratory for Matter Microstructure and Function of Hunan Province, Department of Physics and Synergetic Innovation Center for Quantum Effects and Applications, Hunan Normal University, Changsha 410081, China}
\author{Jia-Shu Niu}
\affiliation{Institute of Theoretical Physics, Shanxi University, Taiyuan, 030006, China}
\author{Zhihui Peng}
\affiliation{Key Laboratory of Low-Dimensional Quantum Structures and Quantum Control of Ministry of Education, Key Laboratory for Matter Microstructure and Function of Hunan Province, Department of Physics and Synergetic Innovation Center for Quantum Effects and Applications, Hunan Normal University, Changsha 410081, China}
\author{Liang Sun}
\affiliation{Institute of Physics, Chinese Academy of Sciences, Beijing, 100190, China}
\author{Zheng Sun}
\affiliation{College of Physics, Sichuan University, Chengdu 610065, China}
\author{Jia Wang}
\affiliation{Institute of Physics, Chinese Academy of Sciences, Beijing, 100190, China}
\author{Puxian Wei}
\affiliation{College of Physics and Optoelectronic Engineering,
Department of Physics, Jinan University, Guangzhou 510632, China}
\author{Lina Wu}
\affiliation{School of Sciences, Xi'an Technological University, Xi'an 710021, P. R. China}
\author{Zhongchen Xiang}
\affiliation{Institute of Physics, Chinese Academy of Sciences, Beijing, 100190, China}
\author{Qiaoli Yang}
\affiliation{College of Physics and Optoelectronic Engineering,
Department of Physics, Jinan University, Guangzhou 510632, China}
\author{Chi Zhang}
\affiliation{Institute of Physics, Chinese Academy of Sciences, Beijing, 100190, China}
\author{Wenxing Zhang}
\affiliation{Tsung-Dao Lee Institute and School of Physics and Astronomy, Shanghai Jiao Tong University, 800 Dongchuan Road, Shanghai 200240, China}
\author{Xin Zhang}
\affiliation{National Astronomical Observatories, Chinese Academy of Sciences, 20A, Datun Road, Chaoyang District, Beijing 100101, China}
\affiliation{School of Astronomy and Space Science, University of Chinese Academy of Sciences, Beijing 100049, China}
\author{Dongning Zheng}
\affiliation{Institute of Physics, Chinese Academy of Sciences, Beijing, 100190, China}
\author{Ruifeng Zheng}
\affiliation{College of Physics and Optoelectronic Engineering,
Department of Physics, Jinan University, Guangzhou 510632, China}
\author{Jian-yong Zhou}
\affiliation{Key Laboratory of Low-Dimensional Quantum Structures and Quantum Control of Ministry of Education, Key Laboratory for Matter Microstructure and Function of Hunan Province, Department of Physics and Synergetic Innovation Center for Quantum Effects and Applications, Hunan Normal University, Changsha 410081, China}
\medskip

\begin{abstract}
Possible light bosonic dark matter interactions with the Standard Model photon have been searched by microwave resonant cavities. In this paper, we demonstrate the cryogenic readout system calibration of a 7.138 GHz copper cavity with a loaded quality factor $Q_l=10^4$, operated at 22 mK temperature based on a dilution refrigerator. Our readout system consists of High Electron Mobility Transistors as cryogenic amplifiers at 4 K, plus room-temperature amplifiers and a spectrum analyzer for signal power detection. We test the system with a superconducting two-level system as a single-photon source in the microwave frequency regime and report an overall 95.6 dB system gain and -71.4 dB attenuation in the cavity's input channel. The effective noise temperature of the measurement system is 7.5 K.
\end{abstract}

\maketitle

\section{Introduction}
\label{sect:intro}

Astrophysical observations indicate the existence of cold and collisionless dark matter (DM) that makes up about 85\% of the matter in our Universe~\cite{Bertone:2004pz,Planck:2018vyg}. While these evidences mostly derive from gravitational effects, a large number of well-motivated particle dark matter candidates have been proposed and extensively searched for, see Ref.~\cite{Boddy:2022knd} for a recent review.
A popular scenario assumes dark matter to take the form of a low-mass bosonic field. The tiny mass of these bosons typically allows a macroscopic de Broglie wavelength, below which the field behaves coherently, well-motivated such candidates include the QCD axion~\cite{Peccei:1977hh,Peccei:1977ur} and axion-like particles, dark photon~\cite{Essig:2013lka}, etc. Their abundance can arise from the so-called `misalignment mechanism'~\cite{Preskill:1982cy,Abbott:1982af,Dine:1982ah,Sikivie:1982qv} in the early Universe, and the preferred boson mass and coupling strength to the Standard Model (SM) are predicted by the specific model. For the QCD axion, a natural mass window for axion being the dark matter is ${\cal O}(10^{-5}- 10^{-3})$ eV~\cite{Sikivie:2006ni}, which corresponds to a converted signal at the microwave energy scale. For axion-like particles and dark photons, wider mass ranges are generally possible. 

The light bosonic dark matter is often motivated by physics at a high energy scale. Their couplings to the SM can be highly suppressed, often making their direct-production search at high-energy experiments difficult. As a result, a class of `haloscope' experiments have been built on Sikivie's resonant cavity design~\cite{Sikivie:1983ip}, in which the cosmic dark matter bosons resonantly convert into an electromagnetic signal in a high $Q$-factor cavity in case it couples to or mixes with the SM photon. In such an experiment, the conversion rate of dark matter is enhanced by the coherence of the light boson ﬁeld and the cavity resonant mode.
Both classical~\cite{Sikivie:1983ip} and quantum-level~\cite{Yang:2022uil} calculations showed that the high-$Q$ factor provides a major boost of the signal rate in the frequency range that matches the energy dispersion of the dark matter field, enabling a high sensitivity narrow-band search. Currently, a large number of proposed and ongoing haloscope experiments ~\cite{DePanfilis:1987dk,Hagmann:1990tj,Kahn:2016aff,Caldwell:2016dcw,HAYSTAC:2018rwy,Marsh:2018dlj,Schutte-Engel:2021bqm,ADMX:2021nhd} are actively searching for dark matter axions and dark photons, and interestingly, with potential applications to high-frequency gravity searches~\cite{Pegoraro:1977uv,Caves:1979kq}.

\begin{figure}[t]
    \centering
    \includegraphics[trim={9cm 0 2cm 0},clip,width=0.7\textwidth]{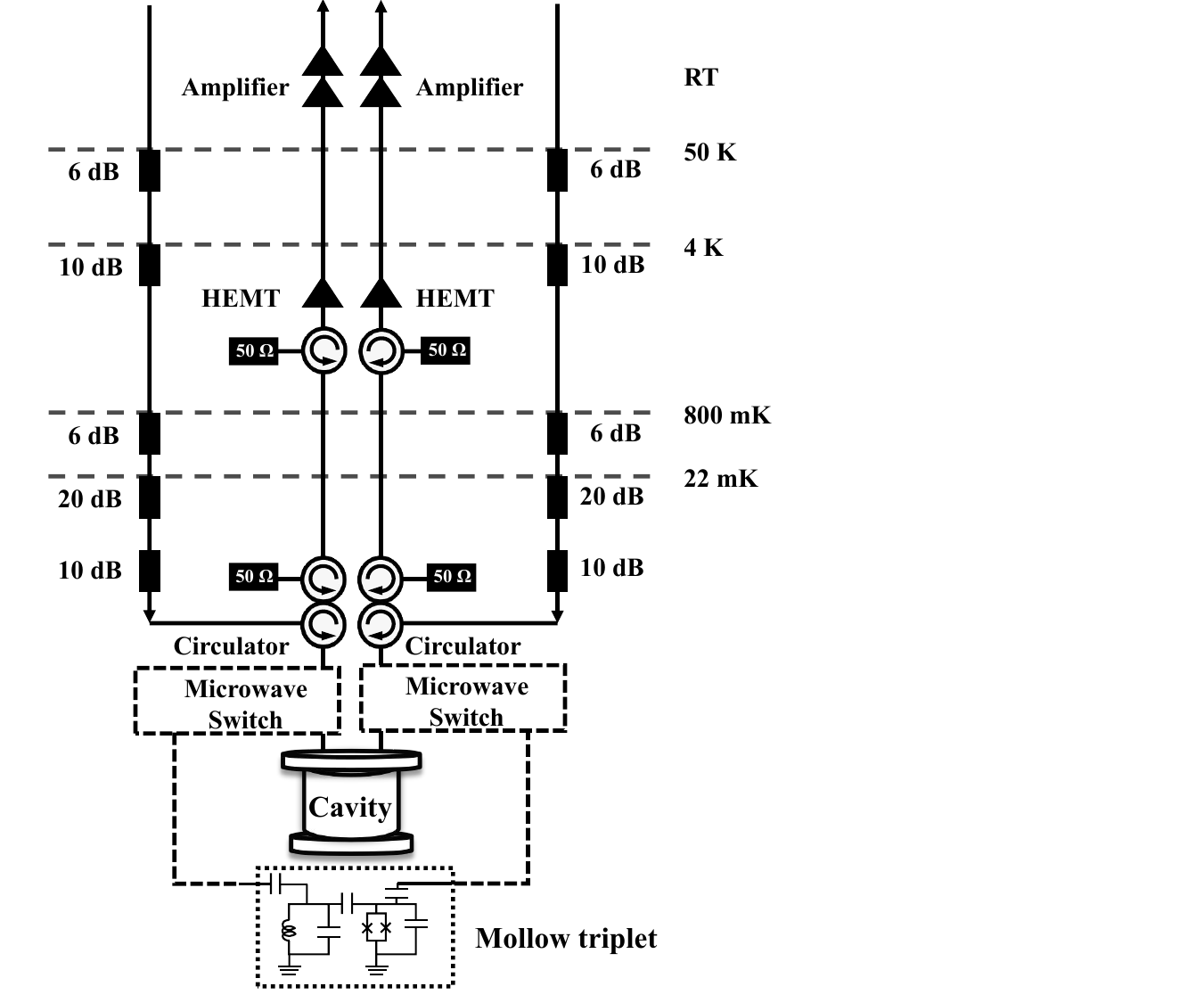}
    \caption{Diagram of the measuring setup for the resonant haloscope cavity. We propose an in situ measuring setup calibration method based on the Mollow triplet of a superconducting two-level system. The two two-position microwave switches regulate either conducting the dark matter search or system calibration.}
    \label{fig:layout}
\end{figure}

The extra Lagrangian terms for the dark photon $A_d$ dark matter are
\be
\Delta L \supset -\frac{1}{4}(F_d^{\mu \nu}F_{d\mu \nu} - 2\chi F^{\mu \nu}F_{d\mu \nu} - 2 m_A^{2} A_d^{2}),
  \label{eq: lagrangian}
\ee
where $F$ and $F_d$ respectively denote the field strengths of the Standard Model photon and dark photon, $m_A$ is a small but nonzero dark photon mass, and $\chi$ is the kinetic mixing parameter~\cite{Holdom:1985ag}. The $\chi\neq 0$ term mixes the dark photon to the Standard Model photon, enabling the dark matter to incite a photon signal in a haloscope cavity with the signal power~\cite{Cervantes:2022epl}
\begin{equation}
    P_{\rm sig.} = \frac{\beta \eta\chi^2 m_A \rho_{\rm DM} V_{\rm eff}}{\beta+1} L(f, f_0, \ql)\,,\quad
\label{eq: power}
\end{equation}
where $\beta\sim 0.95$ is the cavity coupling constant,
$\eta$ is an attenuation factor, $\rho_{\rm DM}$ is the local dark matter density, the dark photon's mass $m_A$ is assumed to match the resonant frequency $f_0$ of the cavity and receives a resonant enhancement from the cavity's loaded quality factor $Q_l$, and $V_{\rm eff}$ is the cavity's effective volume,
\begin{equation}
    V_{\rm eff} \equiv \frac{\left| \int dV {\bf E}(\vec{x}) \cdot {\bf E_d}(\vec{x})\right| ^2}{\int dV |{\bf E}(\vec{x})|^2|{\bf E_d}(\vec{x})|^2}\,,
    \label{eq: veff}
\end{equation}
which depends on the overlap between the dark electric field ${\bf {E}_d}$ and the cavity's resonant mode ${\bf E}$.
$L$ is a Lorentzian detuning factor that describes any off-resonance frequency response in case the measured frequency $f$ differs from $f_0$. For our calibration purposes in this paper, the crucial cavity parameters include the geometric design, resonance frequency $f_0$, and quality factor $Q_l$.

A cavity haloscope typically looks for narrow-width signals above a thermal noise floor. The relevant sources include the thermal noise of the cavity itself, the added Johnson noise from the amplifier/receiver chain, plus quantum fluctuations. For a cavity that operates under the Rayleigh-Jeans limit ($k_b T_{\rm sys} \gg hf$), its noise power per bandwidth is
\begin{equation}
    P_n/\Delta f \simeq G k_b T_{\rm cav.} + k_b T_{\rm D}\,,
    \label{eq: noise power}
\end{equation}
where $k_b$ and $h$ are respectively the Boltzmann and Planck constants, $G$ is the system gain, $T_{\rm cav.}$ is the cavity noise temperature, and $T_{\rm D}$ is the input noise temperature from the amplifier chain. Any signal converted from the dark matter will be picked up with cryogenic amplifiers. To achieve a good signal-to-noise ratio, the measurement must be conducted at a low temperature to reduce thermal noises from both the cavity and the amplifiers.

Modern dilution fridges can maintain temperatures below 50 mK, significantly lower than the quantum noise temperature determined by the microwave frequency range. In the haloscope setup, the primary source of noise contamination often arises from the amplifier input noises ($T_D$ dominated). Commonly used cryogenic amplifiers such as commercial high electron mobility transistors (HEMT)~\cite{HEMT} at 4 Kelvin can provide good linear amplification with added noise around ${\cal O}(10)$ photons. Josephson Parametric Amplifiers (JPA)~\cite{Yamamoto:2008,Lin:2014} or Travelling Wave Parametric Amplifiers (TWPA)~\cite{Macklin:2015} can approach the standard quantum limit in principle and offer higher expected sensitivity.
For the haloscope measurement, critical procedures for estimating the sensitivity also include calibrating the attenuation of input lines, the gain in the amplification channels inside dilution refrigerators, and meticulously measuring the system's noise level. $T_{\rm cav.}$ is the base temperature of dilution refrigerator, which is typically below $50\,$mK. $T_{\rm D}$ is determined by the noise temperature of preamplifiers used in the experiment and dominate the noise temperature in the whole amplification channels. Conducting weak-signal, high-frequency measurements in an ultra-low-temperature environment presents numerous technical challenges. For example, it is often to calibrate the attenuation or gain of the measurement system based on a qubit-cavity coupled system via ac Stark shift in superconducting quantum information experiments~\cite{Schuster:2005}. However, this method is not suitable for the tunable resonant haloscope cavity experiment because the calibrated parameters are only valid at the fixed cavity frequency. With the development of superconducting quantum technologies, the calibration of a measurement system with broadband frequency range based on a superconducting two-level system or single-photon source has been realized~\cite{Hoenigl-Decrinis:2020}.

In this paper, we present the details of essential calibration procedures for the experimental setup of a cryogenic haloscope with a $7.138\,$GHz resonance cavity. We describe the experimental setup in Section~\ref{sect:setup} and discuss the calibration of the resonance cavity and measurement system. We perform test measurements with injected signals, Section~\ref{sect:noise} presents the calibration procedure and gives the effective noise temperature of the measurement system. Then we summarized our results in Section~\ref{sect:summary}.

\section{Experimental setup}
\label{sect:setup}

We conducted the measurements on a cavity device mounted at a $22\,$mK base temperature stage provided by a dilution refrigerator (Model: Bluefors LD 400). The experimental layout is illustrated in Fig.~\ref{fig:layout}. The cavity is located in the $22\,$mK stage, and it contains two identical output ports\footnote{for potential cross-power measurement~\cite{Yang:2022uil} } that connect to two detection chains, and each chain includes a signal input channel and a signal amplification channel. As the target signals are very weak, typically ranging from $-100\,$dBm to $-140\,$dBm, the most effective way to suppress the thermal noise is to attenuate the signal with the noise together in input channels. To reduce the leaked thermal noise from room temperature ($\sim300\,$K) to $4\,$K stage, we mount an attenuator with $6\,$dB attenuation at $50\,$K stage and another attenuator with $10\,$dB attenuation at $4\,$K stage. Furthermore, to suppress the thermal noise arising from temperatures higher than the base temperature, we mount an attenuator with $6\,$dB attenuation at the Still stage ($\sim800\,$mK) and two attenuators with a total $30\,$dB attenuation at base temperature, respectively.

In the amplification channel, the weak output signal from the cavity is first amplified by a cryogenic HEMT amplifier (Model: $\rm{LNF{\textendash}LNC03\_14A}$) at $4\,$K with a typical noise temperature $T_n=4.2\,$K and a gain ($G_{4K})$ at 36 dB, and followed by two amplifiers at room temperature with a total gain $G_{RT}=60\,$dB at room temperature. Note that the effective noise temperature of the haloscope measurement system is determined by the noise temperature of the first amplifier and the loss between the cavity output port and the amplifier. Therefore, it would be very helpful to mount a JPA or TWPA, when available, at base temperature as the first amplifier to enhance the sensitivity. To protect the cavity from noise and reflections from the output side, e.g. noise from the cryogenic amplifier, we mount an isolator ( circulator Model $\rm{LNF{\textendash}CIC4\_12A}$ with one of the three ports terminated with a $50\,\Omega$ cryogenic terminator) with $20\,$dB isolation at $4\,$K stage and an isolator with $20\,$dB isolation at base temperature, respectively. A three-port circulator is mounted at base temperature and it connects the input channel, the cavity port and the output channel to conduct the reflection measurement of the cavity.

\begin{figure}
    \centering
    \includegraphics[width=0.48\textwidth]{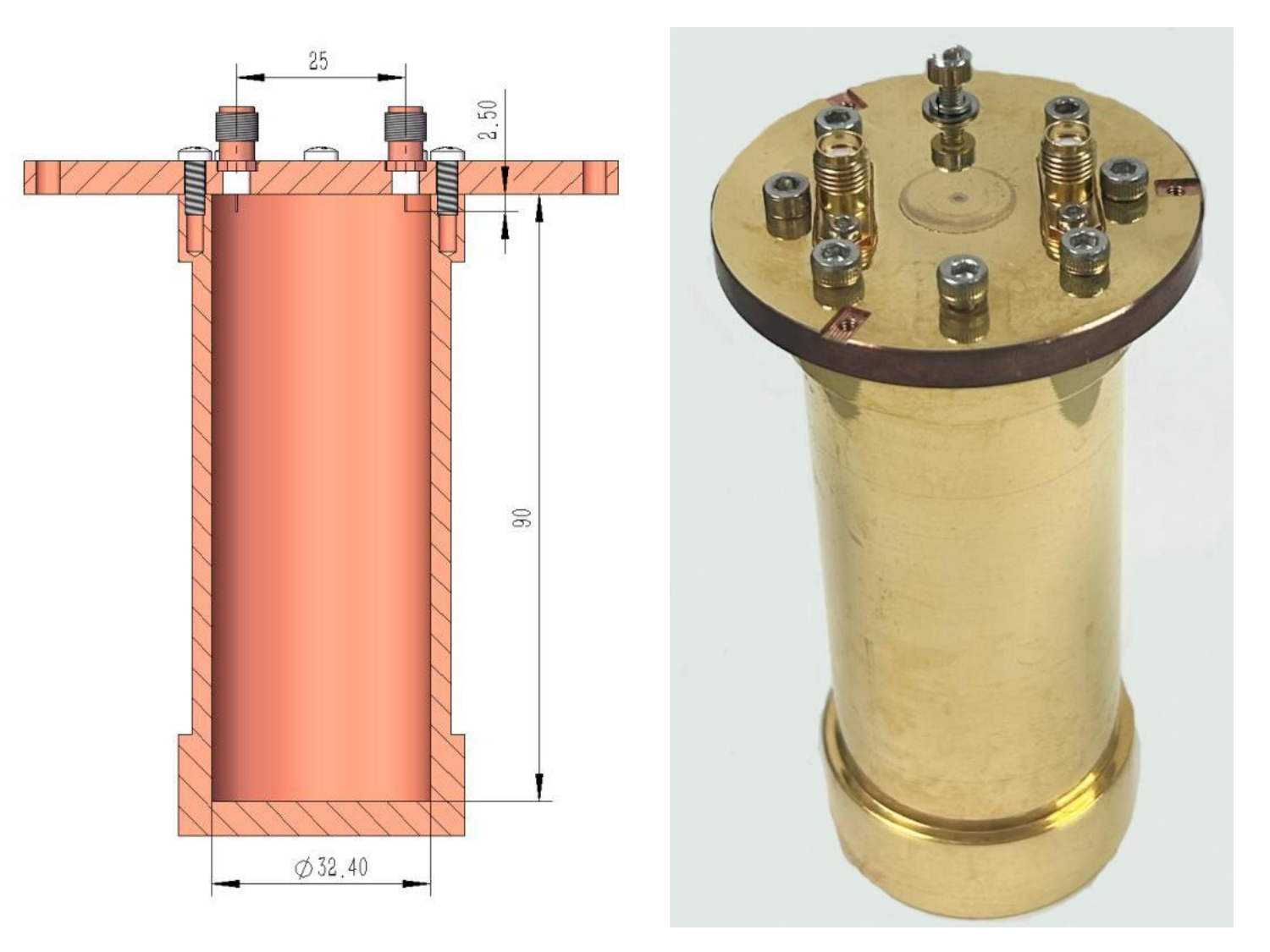}
    \caption{Cross-section design (left) and the exterior (right) of the manufactured resonant cavity.}
    \label{fig:cavity_exterior}
\end{figure}

There is additional attenuation at the coaxial cables as well as the insertion loss from different microwave components, e.g. at the circulators and cavity output ports. The gain of cryogenic and room-temperature amplifiers may change at different signal frequencies. Therefore, we only can estimate the total attenuation and gain for the input and output channels, respectively. To solve these difficulties, we adopt a method based on a superconducting two-level system~\cite{Hoenigl-Decrinis:2020} to carefully calibrate our measurement system with a broadband frequency range, which we will explain further in Sect.~\ref{sect:noise}.

\begin{figure}[t]
    \centering
    \includegraphics[width=0.48\textwidth]{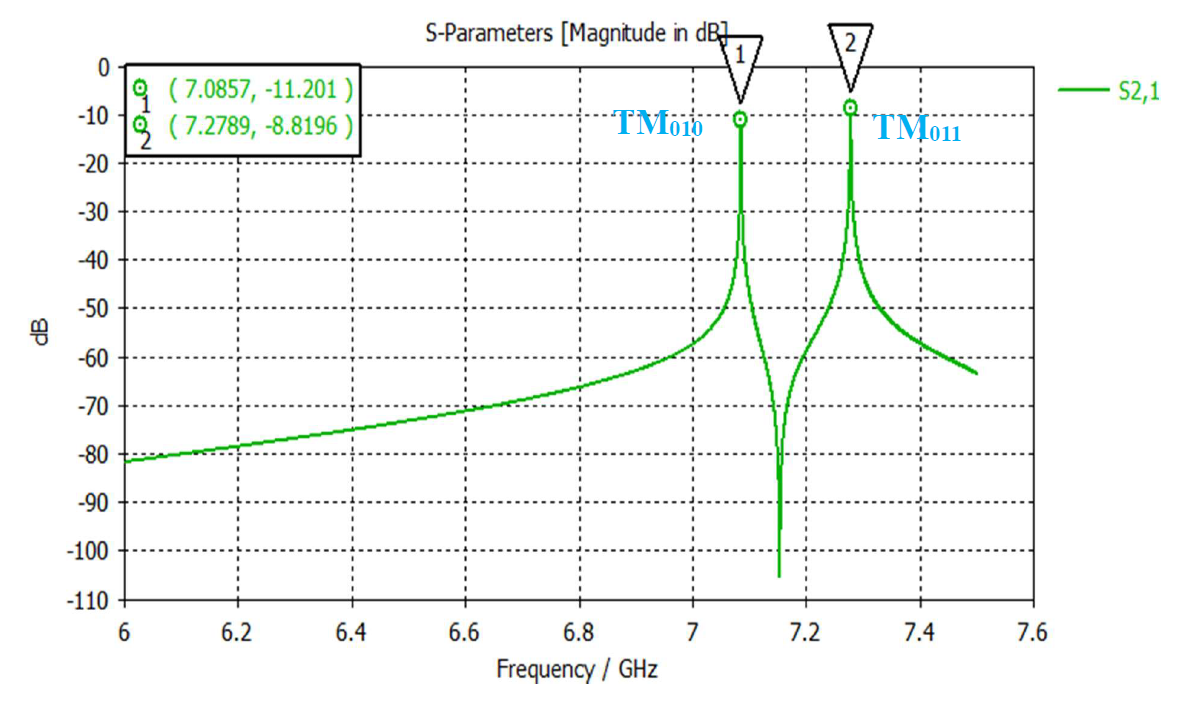}
    \caption{Simulated cavity transmission spectrum containing the $\rm{TM_{010}}$ and $\rm{TM_{011}}$ modes.}
    \label{fig:modes_sim}
\end{figure}

As illustrated in Fig.~\ref{fig:cavity_exterior}, the cylindrical cavity is constructed from pure copper, with an inner diameter of 32.4 mm, a height of 90 mm, and a thickness of 3 mm. It contains a cylindrical inner resonance volume of $74166\,\rm{mm^{3}}$. The lower plane of the cavity is manufactured together with the cavity wall to reduce gaps and losses. The upper surface is tightly connected to the cavity body with 8 screws. The inner surface of the cavity is polished to enhance the quality factor of the cavity. Finally, the cavity surface is coated with gold to prevent oxidization. At both ends of the cavity, the cavity wall is appropriately thickened to improve the mechanical strength of the cavity.

Two ports (KFD 98 SMA connector) are mounted on the upper surface of the cavity. The connector pins are inside the cavity for microwave signal transmission and detection. The width and thickness of the pins are $0.5\,$mm and $0.2\,$mm, and their length inside the cavity is $2.5\,$mm, which determines the loaded quality factor of the cavity. The four lowest resonance frequencies of the cavity are calculated to be $5.6762\,$GHz ($\rm{TE_{111}}$), $7.0853\,$GHz ($\rm{TM_{010}}$), $7.2787\,$GHz ($\rm{TM_{011}}$), and $7.8303\,$GHz ($\rm{TM_{012}}$), respectively. As shown in Fig.~\ref{fig:modes_sim}, we simulate the resonance frequencies around the $\rm{TM_{010}}$ of the cavity with CST software, with the copper conductivity at $5.96\times10^7\,$S/m. In the frequency range from $6\,$GHz to $7.5\,$GHz, the simulated frequencies for $\rm{TM_{010}}$ and $\rm{TM_{011}}$ are $7.0857\,$GHz and $7.2789\,$GHz, respectively, which agree well with our calculations. We $\rm{TM_{010}}$ as our working mode of the cavity because it has a higher form factor in dark matter conversion.
\begin{figure}[h]
    \centering
    \includegraphics[width=0.45\textwidth]{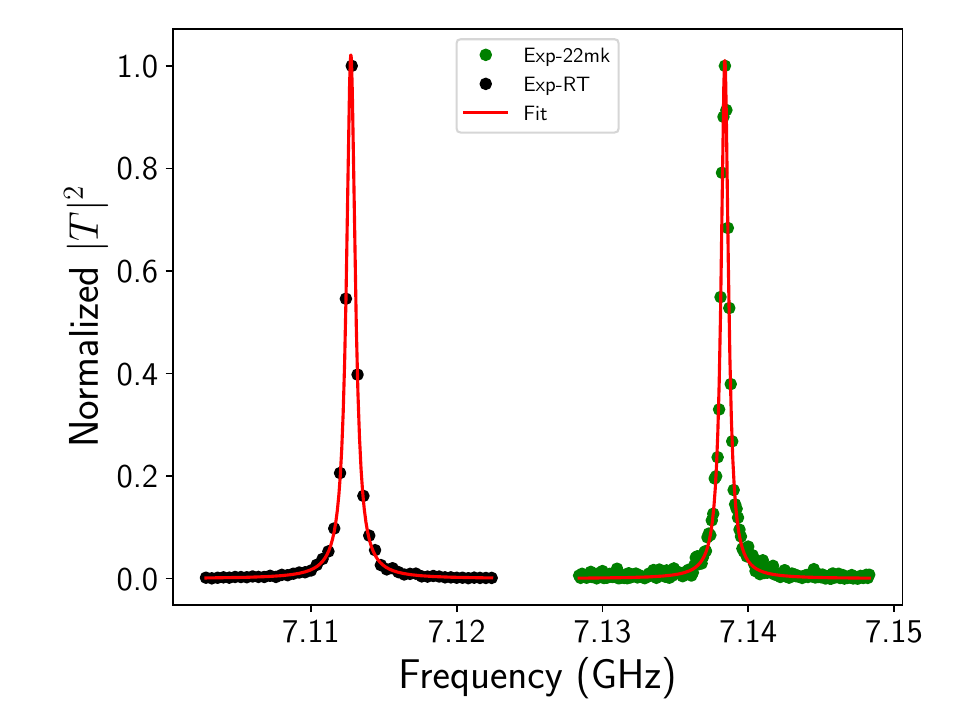}
    \includegraphics[width=0.45\textwidth]{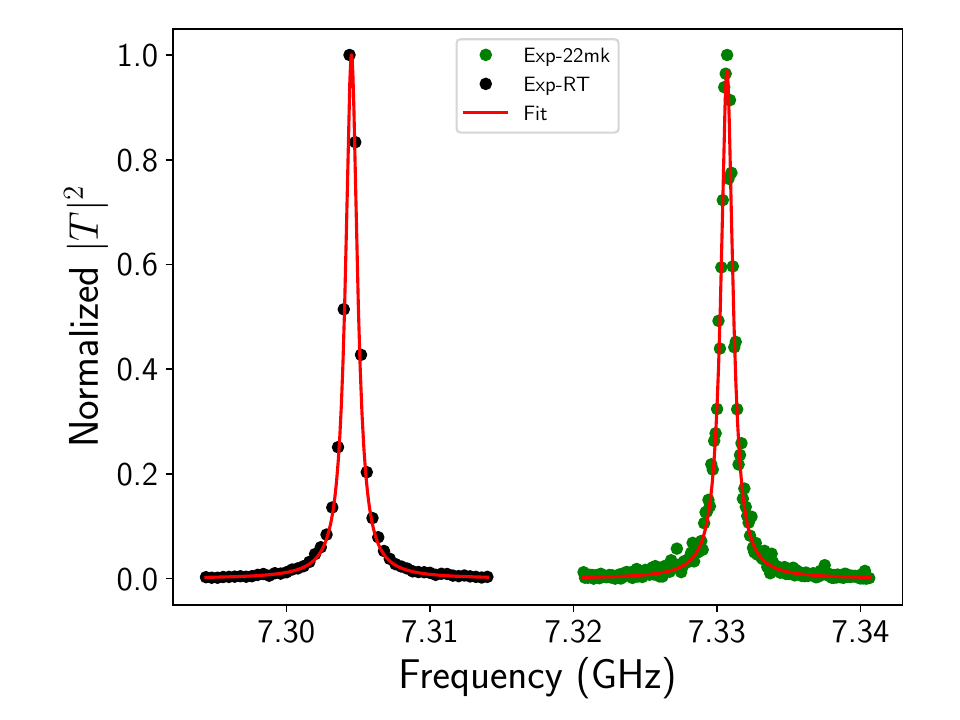}
    \includegraphics[width=0.45\textwidth]{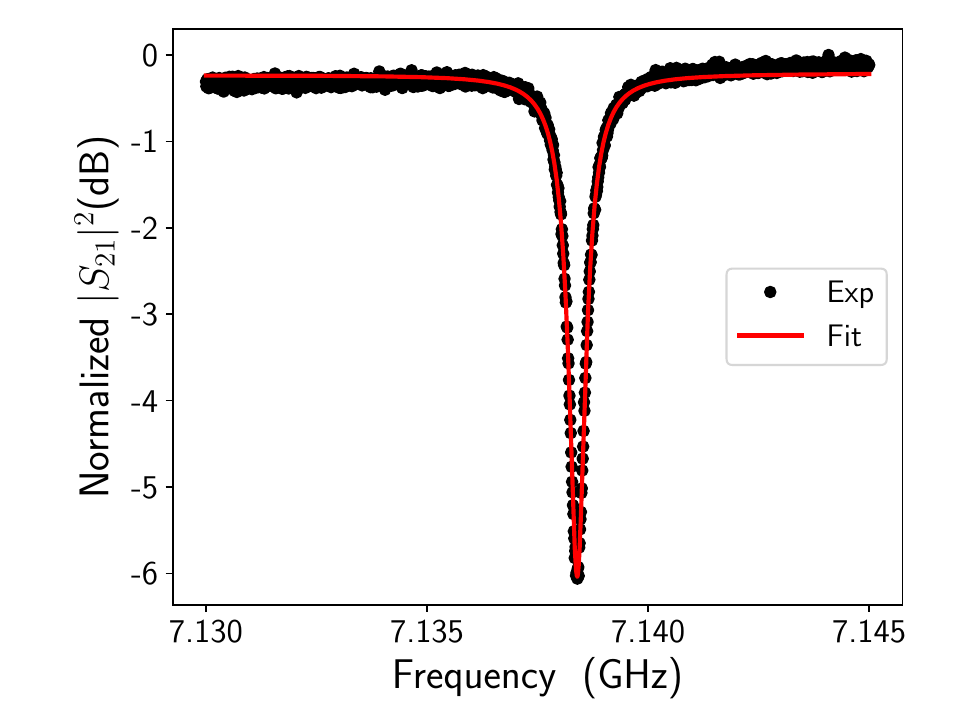}
    \caption{Test results for the $\rm{TM_{010}}$ (upper) and $\rm{TM_{011}}$ (mid) modes.
   The lower panel is the reflection measurement of $\rm{TM_{010}}$ and the fitting result with Eq.~(\ref{qiqc})}
    \label{fig:TM010_tests}
\end{figure}

We performed transmission spectrum measurement on the cavity via a vector network analyzer (VNA) by injecting a probe signal through the input channel on one of the cavity's ports, and then detecting the amplified signal through the output channel on the other port. As shown in the upper and lower panels in Fig.~\ref{fig:TM010_tests}, the $S_{21}$ spectra of the $\rm{TM_{010}}$ and $\rm{TM_{011}}$ modes are presented with an IF frequency $30\,$Hz. To extract the decay rate of the resonant cavity modes, we injected a signal power $P = -145\,$dBm which keeps the cavity’s average photon number  $\langle n\rangle=P/2\hbar\omega_c\kappa$  less than 1 for the measurement of the two cavity modes.
Then we fitted the $S_{21}$ spectra with Lorentzian curves for both modes and extracted resonant frequencies and decay rates. For $\rm{TM_{010}}$ mode, the room-temperature resonant frequency is found to be $\omega_c/2\pi=7.113\,$GHz, with its decay rate $\kappa/2\pi=0.72\,$MHz and quality factor $Q_l=9879$. Therefore, its experimental measurement results (7.113\,GHz) correspond very well with its theoretical predictions (7.0853\,GHz) and simulation outcomes (7.0857\,GHz).  The low-temperature (22 mK) $\rm{TM_{010}}$ frequency is $\omega_c/2\pi=7.138\,$GHz, with $\kappa/2\pi=0.60\,$MHz and $Q_l=11897$ at $22\,$mK. We also measured the $\rm{TM_{011}}$ mode, and the measured resonant frequency is $\omega_c/2\pi=7.305\,$GHz, decay rate $\kappa/2\pi=0.99\,$MHz, quality factor $Q_l=7379$ at room temperature, and $\omega_c/2\pi=7.331\,$GHz, $\kappa/2\pi=0.99\,$MHz, and $Q_l=7405$ at $22\,$mK, respectively.

To extract the extrinsic quality factor, we also measured the reflection ($S_{21}$) of the cavity via the three-port circulator in base temperature and fit it by~\cite{10.1063/1.4919761}
\begin{equation}
\begin{aligned}
&S_{21}  = A \left(1+\alpha \frac{\omega-\omega_c}{\omega_c} \right)\\
&\times \left(1- \frac{\frac{Q_l}{|Q_e|}e^{i\theta} }{1+2iQ_l \frac{\omega-\omega_c}{\omega_c}} \right)
        e^{i (\phi_v \omega + \phi_0)},
\end{aligned}
\label{qiqc}
\end{equation}
where $A$ is the background amplitude, $\omega-\omega_c$ corresponds to the detuning of the detected signal and the cavity frequency, $Q_{l}$ denotes the loaded quality factor, $Q_{e}$ represents the magnitude of extrinsic quality factor, $\theta$ signifies the phase of extrinsic quality factor, $\phi_v$ indicates the phase to accommodate propagation delay to sample, $\phi_0$ describes the phase to accommodate propagation delay from sample, and $\alpha$ characterizes the slope of the signal around the resonance.
The fitting results are depicted in the lower plot of Fig.~\ref{fig:TM010_tests}, revealing $Q_{l}=11006$
and $Q_{e}=22544$. Note that the $Q_{l}$ value obtained by reflection spectrum fitting here is slightly different from the $Q_{l}=11897$ obtained by transmission spectrum fitting above. Both modes experience $\Delta \omega_c/2\pi\approx +25\,$MHz frequency shift after cooling from room to base temperature. Also, the quality factors of both cavity modes at $22\,$mK are higher than those at room temperature due to a lower copper resistivity at $22\,$mK.

\section{Calibration of the measurement system}
\label{sect:noise}

Under strong resonant driving, an incoherent emission spectrum typically exhibits three peaks arises from a two-level system, known as resonance fluorescence or the Mollow triplet~\cite{Mollow:1969}. The explanation of the Mollow triplet can be provided through the dressed-state picture of the driven two-level system~\cite{Mollow:1969,Hoenigl-Decrinis:2020,Peng:2022,Astafiev:2010}. In this picture, the two-level system is dressed by the driving field, leading to the formation of new eigenstates called dressed states. These dressed states exhibit energy level splittings, which result in four distinct transitions: the central peak corresponds to the transition between the undressed ground and excited states, while the two side peaks correspond to transitions involving the dressed states. Therefore, the area of each side peak rigorously corresponds to the power of single-photon $\hbar\omega\Gamma_1$ with different frequency, respectively.

The distance between the two side peaks of the Mollow triplet is related to the energy splitting between the dressed states, which in turn depends on the strength of the driving field and the properties of the two-level system. The distance and two side peaks can be used to calibrate the attenuation and gain of the measurement system with high precision~\cite{Hoenigl-Decrinis:2020}. By accurately determining this distance, one can calibrate the frequency scale of the measurement system, allowing for precise measurements of signal amplitudes and spectral features. Here, we present the calibration of the measurement system's attenuation, gain, and noise level in a broadband frequency range based on the Mollow triplet of a superconducting two-level system with nominal identical measurement system. We use the same superconducting two-level system sample for the calibration same as in Ref.~\cite{Peng:2022}.

\begin{figure}[t]
    \centering
    \includegraphics[width=0.44\textwidth]{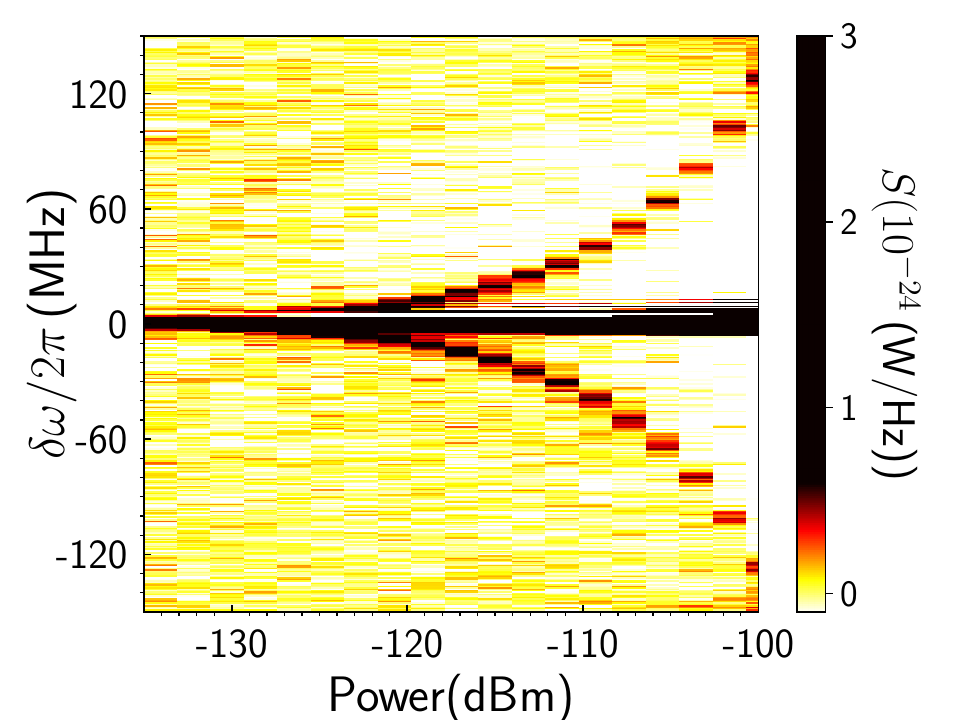}
   \includegraphics[width=0.4\textwidth]{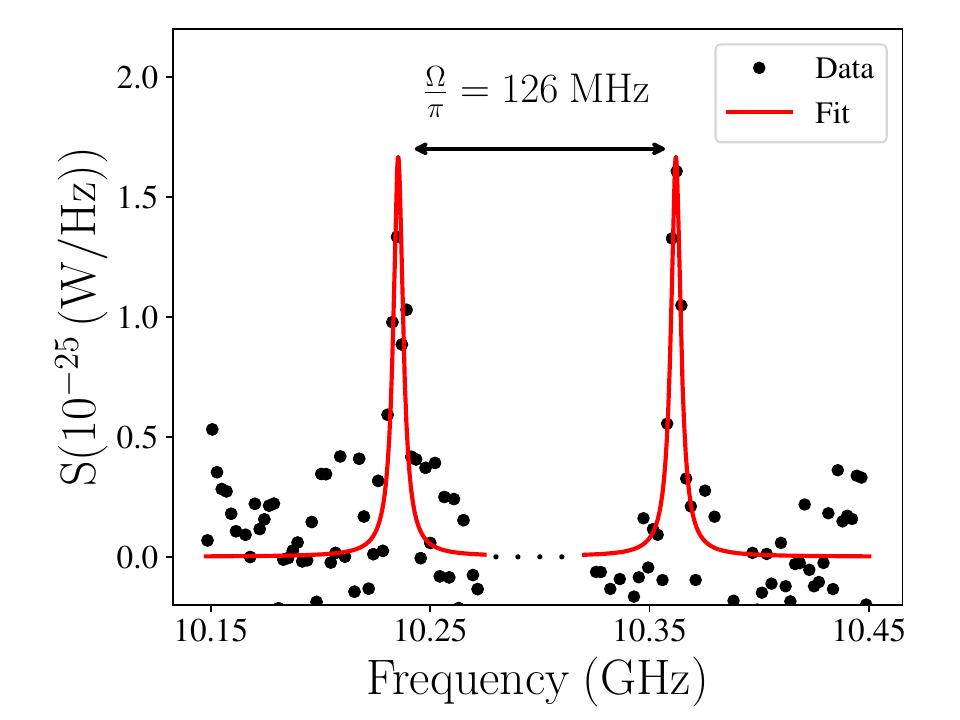}
     \includegraphics[width=0.4\textwidth]{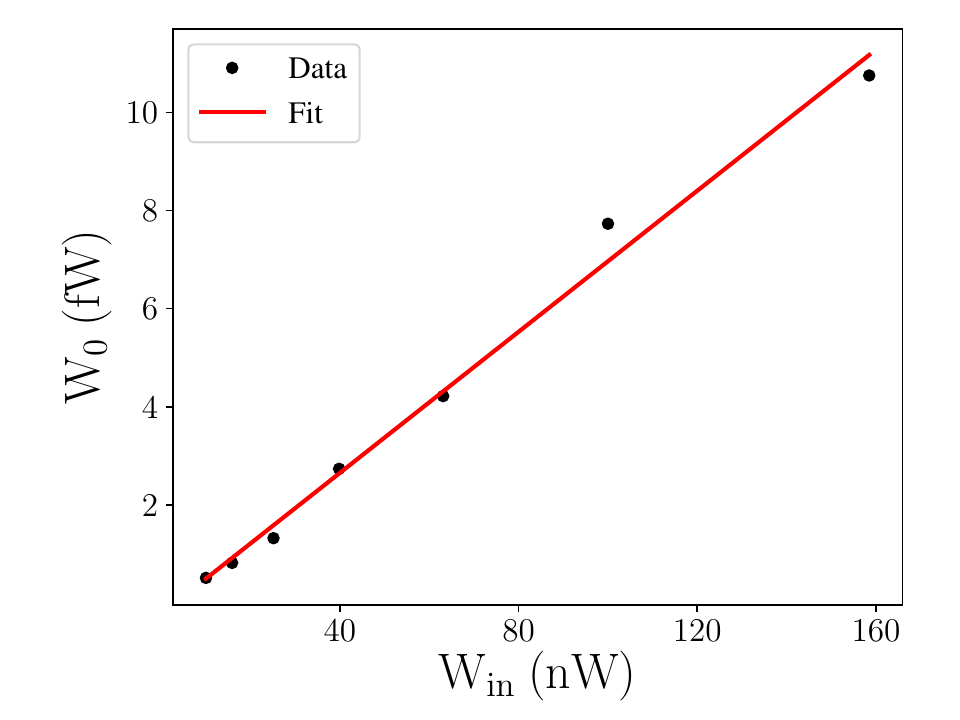}
    \caption{Attenuation calibration in one cavity input channel. The resonance fluorescence spectrum (upper) is shown for the driving power ranging from $-135\,$dBm to $-101\,$dBm with a step of $2\,$dBm. The Mollow triplet's two side peaks (middle) are measured with the driving power at $-107\,$dBm, and the best fit (red) is derived with Eq.~\ref{eq1}. The measured power $W_0$ at the two-level system (lower) is shown versus the input power $W_{in}$ at $10.299\,$GHz, and at room temperature. The attenuation is given by the slope of the linear fit.}
    \label{fig:fluorescence}
\end{figure}

The transition frequency of a superconducting two-level system can be designed in a broad frequency range, i.e. from $1\,$GHz to $12\,$GHz, and tuned by the biased flux through the sample loop in situ. Therefore, we can tune the transition frequency at any frequency that we want to calibrate in this range. For example, we set the two-level system at its maximal transition frequency $\omega_a/2\pi=10.299\,$GHz with smallest dephasing rate and drive it resonantly through an input channel with a coherent microwave field generated by a signal source at room temperature. We measure the Mollow triplet by monitoring the incoherent emission spectrum around $10.299\,$GHz through a spectrum analyzer. The spectral density of the incoherent emission from the two-level system is given by~\cite{Astafiev:2010}

\begin{align} \label{eq1}
    S(\omega)=&\frac{1}{2\pi}\frac{\hbar\omega\Gamma_1}{8}\left(\frac{\Gamma_s}{(\delta\omega+\Omega)^2+\Gamma_s^2}\right. \nonumber\\
        &+\frac{2\Gamma_2}{\delta\omega^2+\Gamma_2^2}+\left.\frac{\Gamma_s}{(\delta\omega-\Omega)^2+\Gamma_s^2}\right),
\end{align}
where $\Gamma_1/2\pi=4.3\pm0.7\,$MHz is the relaxation rate, $\Gamma_2=\Gamma_1/2$ is dephasing rate of the two-level system, respectively, and $\Gamma_s=(\Gamma_1+\Gamma_2)/2$ is the linewidth of side peaks. $\delta\omega$ is the detuning from the transition frequency and $\Omega$ is the Rabi frequency of the two-level system under the resonant driving, respectively.

The intensity plot of the Mollow triplet depends on the driving power on-chip (after calibration) is shown in the upper of Fig.~\ref{fig:fluorescence}. In the middle of Fig.~\ref{fig:fluorescence}, we show the two side peaks in the emission spectrum when the driving power on-chip is around $-107\,$dBm. The distance between the two side peaks is $2\Omega=2\pi\times126\,$MHz by fitting the resonance fluorescence spectrum with Eq.~(\ref{eq1}). We calculate the power sensed by the two-level system according to $W_0= \hbar\omega\Omega^2/(2\Gamma_1)$. Then, we can plot $W_{\rm in}$ versus $W_0$ with a linear fit and the attenuation is $-71.4\pm0.2\,$dB in the input channel which is given by the slope. We can get quite high calibration precision because the fit can give a low uncertainty $\Delta\Omega/\Omega\leq0.01$, where $\Delta\Omega$ is the uncertainty of $\Omega$. The gain of the output line can also be obtained by the Mollow triplet measured by the spectrum analyzer at room temperature. By fitting the two side peaks with Eq.~\ref{eq1}, we get the gain of the output channel of our measurement system is $95.6\pm1.0\,$dB. With the calibrated attenuation and gain in our measurement, we can calibrate the emission spectrum measured by a spectrum analyzer. The measured noise spectral density of the HEMT amplifier placed at $4\,$K stage is around $1.03\times10^{-22}\,$W/Hz. According to $2\pi S(\omega)=k_{\rm B}T_{\rm D}$~\cite{Astafiev:2010}, we can deduce the effective noise temperature of the measurement is around $7.5\,$K with an accuracy about $\pm1\,$dB and the uncertainty is determined by the uncertainty of calibrated gain. We calibrated the attenuation, gain, and effective noise temperature at the maximal transition frequency (10.299 GHz) of the two level system. Due to the frequency difference,
there can be some extra uncertainty in the cable attenuation, connector loss and the gain from HEMT and room amplifiers at 7.138 GHz. The nominal attenuation from microwave cables at $7.138\,$GHz is around $2\,$dB smaller than that at $10.299\,$GHz. The nominal gain from HEMT at $7.138\,$GHz is $2\,$dB smaller. There is no nominal difference from the gain of room temperature amplifiers between $7.138\,$GHz and $10.299\,$GHz. The frequency dependence of total loss from all connectors is more complicated, but the total loss should be smaller than $2\,$dB. Therefore, we can roughly estimate the attenuation is $-71.4\pm3\,$dB, the gain is $95.6\pm3\,$dB, and the effective noise temperature is $7.5\,$K with an accuracy around $\pm3\,$dB for the measurement system at $7.138\,$GHz. This uncertainty from frequency difference can be further removed by follow-up in situ measurements.

\section{Summary}
\label{sect:summary}

In summary, we have built a resonant haloscope cavity measurement system based on a dilution refrigerator and tested the cavity and measurement system at room temperature and 22 mK. The cylindrical two-port copper cavity has a 74166 mm$^3$ volume and it is designed to operate at a 7.138 GHz ${\rm TM_{010}}$ mode with a low-temperature quality factor $Q=10^4$. We calibrate the attenuation of the input channel, the gain in the output channel, and the effective noise temperature of the measurement system based on measuring the Mollow triplet with a superconducting two-level system. The calibrated attenuation is $-71.4\pm3\,$dB, the gain is $95.6\pm3\,$dB, and the effective noise temperature is $7.5\,$K with an accuracy $\pm3\,$dB with our HEMT amplifiers, which contribute the majority of the measured noise. In perspective, quantum-limited preamplifiers, such as JPA and TWPA, will significantly reduce the effective noise temperature in the system. The advantage of our calibration method is the ability for in situ system calibration in a broad frequency range. This method can also be well-suited for frequency-adjustable narrow-band cavity experiments.

\vspace{1cm}

{\bf Acknowledgements \\}
\ \  Measurements were performed at Hunan Normal University in Changsha. This work is supported in part by the Scientific Instrument Developing Project of the Chinese Academy of Sciences (YJKYYQ20190049), the International Partnership Program of Chinese Academy of Sciences for Grand Challenges (112311KYSB20210012), the National Natural Science Foundation of China (No. 12074117, 92365209, 12150010, 11875062, 11947302, 12047503, 12074117, 61833010, 12061131011, 12150410317, 11905149 and 12275333) and the Beijing Natural Science Foundation (IS23025).

\bibliographystyle{unsrt}
\bibliography{refs.bib}

\end{document}